\documentclass[journal,twoside]{IEEEtran}

\usepackage{cite}

\usepackage{graphicx}

\usepackage{amsmath}
\usepackage{amssymb}

\usepackage{array}

\usepackage{siunitx}

\usepackage{hyperref}

\usepackage[acronym]{glossaries}
\glsdisablehyper

\newacronym{dut}{DUT}{Device-Under-Test}
\newacronym{ddsrf}{DDSRF-PLL}{Doubly-Decoupled Synchronous Reference-Frame Phase-Locked-Loop}
\newacronym{wea}{WT}{Wind Turbines}
\newacronym{mhil}{MHiL}{Mechanical Hardware-in-the-Loop}
\newacronym{phil}{PHiL}{Power Hardware-in-the-Loop}
\newacronym{frt}{FRT}{Fault-Ride-Through}
\newacronym{afe}{AFE}{Active-Front End}
\newacronym{fft}{FFT}{Fast-Fourier-Transform}
\newacronym{poc}{POC}{Point-of-connection}

\newcommand{\irms}{I_\mathrm{rms}}
\newcommand{\udc}{V_\mathrm{dc}}
\newcommand{\cdc}{C_\mathrm{dc}}
\newcommand{\fsw}{f_\mathrm{sw}}
\newcommand{\rbrake}{R_\mathrm{B}}
\newcommand{\utb}{U_\mathrm{DC, tb}}

\newcommand{\uSAbc}{v_\mathrm{s,abc}}
\newcommand{\iSAbc}{i_\mathrm{s,abc}}
\newcommand{\isd}{i_\mathrm{sd}}
\newcommand{\isdDot}{\dot{i}_\mathrm{sd}}
\newcommand{\isq}{i_\mathrm{sq}}
\newcommand{\isqDot}{\dot{i}_\mathrm{sq}}
\newcommand{\ird}{i_\mathrm{\mu}}
\newcommand{\irdDot}{\dot{i}_\mathrm{\mu}}
\newcommand{\usd}{v_\mathrm{sd}}
\newcommand{\usq}{v_\mathrm{sq}}
\newcommand{\tauS}{\tau_\mathrm{s}}
\newcommand{\tauR}{\tau_\mathrm{r}}
\newcommand{\sigmaS}{\sigma_\mathrm{s}}
\newcommand{\sigmaR}{\sigma_\mathrm{r}}
\newcommand{\Ls}{L_\mathrm{s}}
\newcommand{\Lr}{L_\mathrm{r}}
\newcommand{\Lh}{L_\mathrm{h}}
\newcommand{\LsigmaS}{L_\mathrm{\sigma s}}
\newcommand{\LsigmaR}{L_\mathrm{\sigma r}}
\newcommand{\Rs}{R_\mathrm{s}}
\newcommand{\Rr}{R_\mathrm{r}}

\newcommand{\omegaRMech}{\omega_\mathrm{gen}}
\newcommand{\omegaSl}{\mu_\mathrm{r}}
\newcommand{\Tmech}{T_\mathrm{gen}}
\newcommand{\SV}{S_\mathrm{V}}

\newcommand{\omdq}{\omega_\mathrm{dq}}

\newcommand{\Td}{T_\mathrm{d}}

\newcommand{\Tcc}{T_\mathrm{cc}}

\newcommand{\DistSpec}{D_\mathrm{spec}}
\newcommand{\mathcomma}{\quad \mathrm{,}}
\newcommand{\mathdot}{\quad \mathrm{.}}

\usepackage{etoolbox}
\makeatletter
\patchcmd{\IEEEbiographynophoto}
  {\vskip 4\baselineskip plus 1fil minus 0\baselineskip}
  {\vskip 1\baselineskip}
  {}{\PackageWarning{main_ieee}{Could not patch IEEEbiographynophoto}}
\makeatother

\begin{document}
%
\title{Grid-Compliance Testing of Wind Turbines Through Virtualized Component Tests: Testing Concept and Test Bench}
%
\author{Simon~Moritz,
        Yoga~Viknesh~Kannan,
        Florian~Andresen,
        Laura~Fuentes~Grau,
        Manuel~Fischer,
        Johannes-Martin~Hördler,
        Sabri~Özer,
        Martin~Brennecke,
        Philipp~Kersting,
        Andreas~Thul,
        Simon~Steentjes,
        Ferdinanda~Ponci,
        Antonello~Monti,
        Rik~W.~De~Doncker,
        Heike~Vallery,
        and~Maximilian~Basler
\thanks{\textbf{Author's version.} This is the author's version of an article published in \textit{atp magazin}, vol.~08/2026, pp.~48--61, 2026, url:~\href{https://atpinfo.de/ausgabe/atp-magazin-82026/}{https://atpinfo.de/ausgabe/atp-magazin-82026/}. The published article may differ from this version. Please cite the published article: S.~Moritz \textit{et~al.}, ``Grid-Compliance Testing of Wind Turbines Through Virtualized Component Tests: Testing Concept and Test Bench,'' \textit{atp magazin}, vol.~08/2026, pp.~48--61, 2026.}
\thanks{This work was supported by the German Federal Ministry of Economic Affairs and Energy (BMWE) under Grant No.~03EE3079A.}
\thanks{S.~Moritz, H.~Vallery, and M.~Basler are with the Institute of Automatic Control and the Center for Wind Power Drives, RWTH Aachen University, Aachen, Germany.}
\thanks{Y.~V.~Kannan, L.~Fuentes~Grau, F.~Ponci, and A.~Monti are with the Institute for Automation of Complex Power Systems and the Center for Wind Power Drives, RWTH Aachen University, Aachen, Germany.}
\thanks{F.~Andresen and R.~W.~De~Doncker are with the Institute for Power Generation and Storage Systems and the Center for Wind Power Drives, RWTH Aachen University, Aachen, Germany.}
\thanks{P.~Kersting, A.~Thul, and S.~Steentjes are with the Institute of Electrical Machines and the Center for Wind Power Drives, RWTH Aachen University, Aachen, Germany.}
\thanks{M.~Fischer, J.-M.~Hördler, and S.~Özer are with dSPACE SE \& Co. KG.}
\thanks{M.~Brennecke is with FGH Zertifizierungsgesellschaft mbH.}}

\markboth{Author's version of an article published in atp magazin, no.~8, 2026. Please cite the published article.}{Moritz \MakeLowercase{\textit{et al.}}: Grid-Compliance Testing of Wind Turbines Through Virtualized Component Tests}

\maketitle

\begin{abstract}
In the course of the energy transition toward renewable energies, the share of systems coupled to the electrical grid via power electronics converters, including wind turbines (WT), is steadily increasing. These systems must be operated in a grid-supportive manner with respect to grid stability and support, and must be tested accordingly in advance. This paper presents a Power Hardware-in-the-Loop (PHiL)-based testing procedure for grid-compliance testing of WT, addressing the conceptual design, modeling, and construction of a corresponding megawatt-class PHiL component test bench. The procedure developed here enables testing of the full-scale converter with its associated control system as the Device-Under-Test (DUT), whereby the electrical behavior of the generator, the mechanical components, and the grid to be fed are replicated by means of power electronics actuators in real-time emulation, which considerably simplifies the mechanical setup. In addition to local operation, the geographically distributed simulation of individual subsystems, namely the aerodynamics via the internet and the grid model via a dedicated fiber-optic link, is considered for the first time in the context of certification-relevant tests. Beyond the component-based testing procedure, both the necessary models at their relevant level of model depth are presented, and the converter-based component test bench developed for the experimental implementation is described in detail. The concept presented here thus advances already established procedures of IEC TS 61400-21-4 and can contribute to the further establishment of PHiL test benches in the wind energy industry. Initial experimental results demonstrate high emulation quality: The grid emulation meets all requirements of relevant standards regarding voltage quality, and the geographically distributed simulation of mechanical components achieves sufficiently high accuracy at a round-trip latency of less than \SI{10}{\milli\second}. The grid-side geographically distributed simulation also yields a sufficiently small latency of \SI{160}{\micro\second}.
\end{abstract}


\IEEEpeerreviewmaketitle

\section{Introduction}
\IEEEPARstart{G}{rid} compliance testing of \gls{wea} constitutes a mandatory prerequisite before connecting them to the public power grid. It ensures that the plant operates in a grid-supportive manner, does not destabilize the grid, and, in certain cases, such as grid faults, actively contributes to grid stability and restoration. The testing covers several categories, ranging from evaluation of active and reactive power feed-in, through assessment of grid interaction effects, to testing under fault conditions, including two-phase and three-phase voltage dips and swells.

The legal framework is based on a series of national and international regulations: First, the minimum requirements for grid connection of generating units are harmonized at the EU level through the \textit{Commission Regulation (EU) 2016/631}~\cite{european_commission_commission_2016} and further specified at the German national level by the \textit{Verordnung zum Nachweis von elektrotechnischen Eigenschaften von Energieanlagen (NELEV)}~\cite{noauthor_verordnung_2017}. For the German power grid, this leads to the application of \textit{VDE-AR-N 4110}~\cite{vde_verband_der_elektrotechnik_elektronik_informationstechnik_ev_technical_2023} for \gls{wea} intended for connection to medium-voltage grids, where the standard form of verification consists of a plant certificate based on unit certificates and simulation models. To this end, the certification procedure according to \textit{FGW TR8}~\cite{fgw_ev_technical_2019} applies, which references the guideline \textit{FGW TR3}~\cite{fgw_ev_technical_2022} for measurement-based verification.

Since field testing, i.e., with the fully assembled and operational system, requires substantial time, incurs correspondingly high costs, and lacks reproducibility, considerable efforts have been and continue to be made to enable testing on dedicated test benches. This led to a classification of test-bench types in the technical specification IEC TS 61400-21-4:2025~\cite{international_electrotechnical_commission_wind_2025}, which specifies the required test-bench type for each test case. Table~\ref{tab:test-bench-after-IEC} summarizes this classification.

\begin{table}[!t]
    \caption{Classification of test benches according to IEC TS 61400-21-4. The ordering reflects descending scope of permissible test cases. For instance, tests that require a 3a-classified test bench may also be conducted on a 1a-classified test bench.}
    \label{tab:test-bench-after-IEC}
    \centering
    \begin{tabular}{c|p{0.8\columnwidth}}
    \hline
        1a & Nacelle test bench (closed-loop testing) with \gls{mhil} control \\
        1b & Nacelle test bench (open-loop testing) \\
        2a & Electrical generation test bench (closed-loop testing) with \gls{mhil} control \\
        2b & Electrical generation test bench (open-loop testing) \\
        3a & Converter test bench (closed-loop testing) with \gls{phil} control \\
        3b & Converter test bench (open-loop testing) \\
        4a & Controller test bench (closed-loop testing) with signal-level HiL \\
        4b & Controller test bench (open-loop testing) \\
        5 & Auxiliary test bench \\
        \hline
    \end{tabular}
\end{table}

Notably, some relevant test categories require at least Class~1 or Class~2 test benches, which necessitate a physically present generator, the converter, and the \gls{wea} control unit (with the exception of Class~2b). However, for Type-IV \gls{wea}, i.e., turbines equipped with a full-scale converter, the converter predominantly determines the electrical characteristics at the \gls{poc}, thereby enabling component testing. The idea is to test the full-scale converter along the \gls{wea} control unit with a \gls{phil} system, where all other components, e.g., generator and grid, are emulated in a real-time system, and load application is achieved purely by power electronics. Additionally, given the large share of simulated components, the question arises of whether some can be simulated in a geographically distributed manner while being coupled to the on-site test bench via a suitable communication infrastructure, which would considerably improve the accessibility of cost-intensive testing infrastructure for manufacturers.

For the wind energy industry, further establishing \gls{phil} test benches for component testing would be highly advantageous during the development phase and for carrying out grid-compliance measurements. First, the mechanical setup requires less complexity, space, and time compared to \gls{mhil}-based nacelle test benches. Second, the \gls{dut}, i.e., the converter and \gls{wea} controller, can be examined in isolation, which considerably simplifies fault diagnosis when tests fail. The geographic distribution also opens the prospect of treating test benches as `Infrastructure-as-a-Service` in the future, thereby protecting manufacturers' intellectual property, since remotely connected models need not be disclosed while their use in testing remains possible. These aspects can therefore pave the way for time- and cost-efficient testing of \gls{wea} with respect to system integration and electrical properties.

In this paper, we propose to capitalize on these insights to establish a new component-level testing concept and integrate it into a megawatt-scale \gls{phil} test bench, which requires, in addition to mechanical component and grid models, modeling the generator in a real-time simulation and applying the load purely through power electronics. Several implemented modeling depths of the generator enable an investigation into the sensitivity of grid-compliance tests on the used model. 

Additionally, we introduce the concept of geographically distributed simulation for grid-compliance testing. This type of test bench can therefore be regarded as an extension of Class~3a in the current classification. We present the conceptual design and construction of such a test bench. The objective of the proposed testing concept and test bench is to experimentally investigate their suitability for grid-compliance testing of \gls{wea}. 

A concept for \gls{phil} testing of \gls{wea} converters based on the Class~3a test-bench classification was presented previously, cf.\ \cite{neshati_grid_nodate}. Our proposed testing concept differentiates itself in two aspects: First, the \gls{phil} system can operate both as a voltage source and as a current source. This is particularly advantageous for generator emulation, since no transformer between the \gls{phil} converter and the \gls{dut} is required, thereby entirely avoiding parasitic effects such as saturation. Emulating the generator with a current source is also more dynamically accurate, as it avoids approximating current derivatives when calculating voltage setpoints. Second, geographically distributed simulation for grid compliance testing of \gls{wea} has not been considered to date, yet it would substantially enhance the flexibility of future test infrastructure. Together, these aspects establish the testbench as a modular, geographically distributed platform for comprehensive PHiL-based type testing of wind turbine converters, thereby enabling flexible and cost-effective grid-compliance validation.

The remainder of this paper is structured as follows: Section~\ref{sec:concept-and-modeling} provides a conceptual overview of the test bench and introduces the employed models. Section~\ref{sec:test-bench-setup} then details the hardware setup of the test bench, and Section~\ref{sec:preliminary-results-and-discussion} presents preliminary results derived from experimental measurements, and a discussion. Finally, Section~\ref{sec:conclusion-and-outlook} draws conclusions and provides an outlook for future work.

\section{Concept and System Overview}
\label{sec:concept-and-modeling}
\subsection{Conceptual overview}
\begin{figure*}[!t]
\centering
\includegraphics[width=\textwidth]{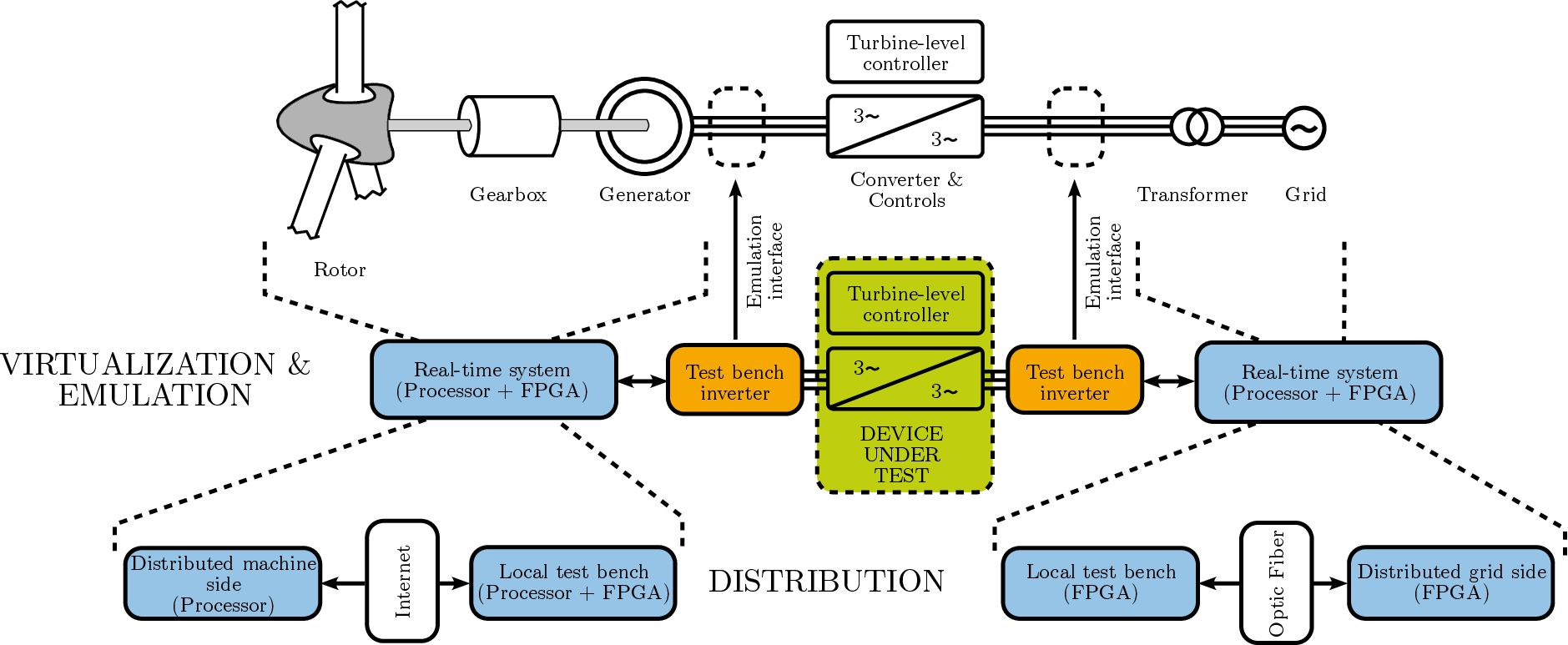}
\caption{Two-stage conceptual diagram of the proposed testing concept.}
\label{fig:testbench-concept}
\end{figure*}

The testing concept presented here tests the full-scale converter, including its associated control system and turbine-level controller, as the \gls{dut}. To obtain meaningful results comparable to field conditions, the behavior of the \gls{dut} at the \gls{poc}, which arises from the overall system behavior of coupled multiphysical systems, including their controls, must be reproduced realistically. To achieve this, the components that are not physically present are modeled, embedded in a multiphysical real-time simulation, and emulated through power electronics load application at the test bench. Together with the geographically distributed simulation of individual subsystems, this yields a two-stage emulation concept, illustrated in Fig.~\ref{fig:testbench-concept}.

From the perspective of the \gls{dut}, two power interfaces arise: First, a power-generating interface that handles the energy conversion chain from the kinetic energy in the wind, through rotational mechanical energy, to electrical energy. This part of the system comprises a wind field model, an aerodynamic rotor model, a drivetrain, and a generator, and is referred to as the \textit{machine side} in the following. Second, a power-feeding interface, which couples the \gls{dut} to the electrical grid. Here, the grid behavior, comprising a voltage source, grid impedance, and transformer, must be emulated. Hence, we refer to this part of the system as the \textit{grid side}. In addition to these two electrical power interfaces, several signal interfaces are required for embedding the \gls{dut} in the test bench. As an essential part of the proposed testing concept, the models employed at the virtualization and distribution levels (cf.\ Fig.~\ref{fig:testbench-concept}) are presented in the following.

\subsection{Reference Simulation: Design and Analysis of the Overall System}
The proposed testing concept involves the virtualization of many components of a \gls{wea}. Moreover, the load application of the \gls{phil} system introduces additional influences in the closed control loop that are absent in an actual \gls{wea}. These primarily include limited bandwidth and time delays, which can critically affect the emulation quality \cite{ren_improve_2007}. To validate the implemented sub-models and ensure high emulation quality despite these additional influences, a reference simulation is required. This reference simulation encompasses both the virtualized components and models of the \gls{dut}, the load application system, and the geographic distribution. By performing comparative simulations with and without load application, we ensure on the one hand that the \gls{phil} system does not degrade the emulation quality, or does so only within an acceptable margin. On the other hand, implementing and validating the sub-models as well as the overall system enables subsequent verification of the implementation on the real-time systems, which is indispensable for the reliability of the results.

\begin{figure*}[!t]
	\centering
	\includegraphics[width=\textwidth]{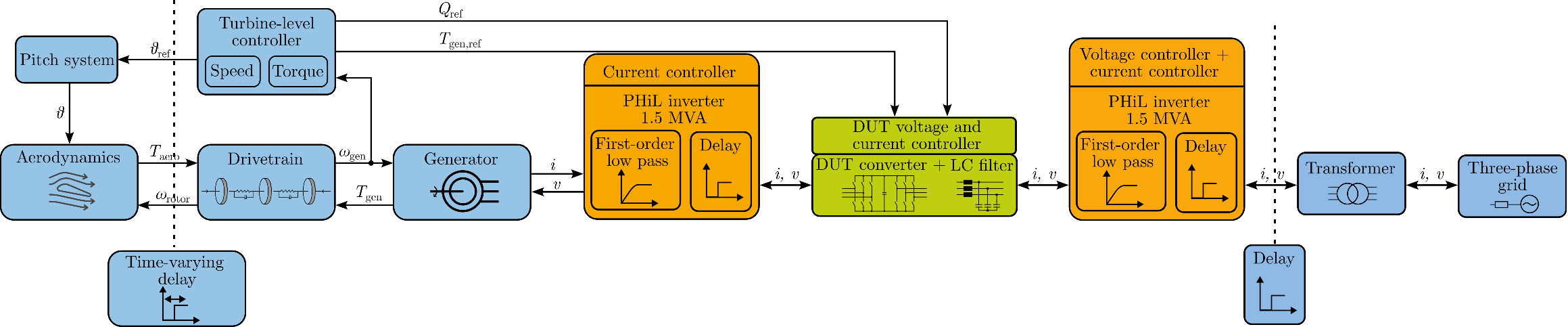}
	\caption{Reference simulation with models and interfaces.}
	\label{fig:reference-simulation-complete}
\end{figure*}

The architecture of the reference simulation is shown in Fig.~\ref{fig:reference-simulation-complete} and follows the conceptual structure of Fig.~\ref{fig:testbench-concept}.

\subsection{Component Models}
\subsubsection{Aerodynamics and Pitch System}
For accurate simulation of the aerodynamic rotor behavior, we employ the software RAISE, which performs an aeroelastic simulation based on the Blade-Element-Momentum (BEM) theory \cite{marnett_real-time_2014}. The mean wind speed, prescribed by the respective test case, is converted into either a laminar or turbulent three-dimensional wind field, depending on the test case. Additionally, the rotor speed and the pitch angle are required as inputs. The simulation yields the instantaneous aerodynamic torque acting on the drivetrain. To approximate the behavior of physical pitch actuators, the pitch system is modeled as a first-order lag element.
\subsubsection{Drivetrain}
According to the applicable standards, the drivetrain is modeled as an $n$-mass oscillator, in which the individual masses are coupled via spring-damper elements \cite{uwe_jassmann_hardware---loop_2018}. By using up to three masses, namely rotor, hub, and generator, the first two eigenmodes of the drivetrain can be captured. Depending on the test case, the three-mass oscillator can also be replaced by a two- or one-mass oscillator. The decision regarding the appropriate level of model detail is made based on the sensitivity analysis presented in \cite{jassmann_certbench_2021}.
\subsubsection{Generator}
A central aspect of the proposed testing concept is the emulation of the generator, implemented at several levels of model fidelity to enable a sensitivity analysis of its influence on test results. We model a squirrel-cage induction machine. Although such a machine is not among the most commonly used in modern wind turbines, it was selected here due to the availability of measurements, which we used for model validation. Since the model runs purely in software, it poses no limitation to the proposed testing concept, as other types of electrical machines suitable for full-converter operation can readily be implemented and used on the test bench.

The generator models take the three-phase generator voltage $\uSAbc$ and the mechanical speed $\omegaRMech$ as inputs, and produce the three-phase generator current $\iSAbc$ and the mechanical torque $\Tmech$ as outputs. The dynamic modeling follows a rotor-flux-oriented formulation in the dq reference frame rotating at the angular frequency $\omdq$, obtained through the amplitude-invariant Park transformation \cite{orourke_geometric_2019}.
The first level of model fidelity describes a classical rotor-flux-oriented fundamental-wave model, formulated according to \cite{gerling_electrical_2015}, which is described by
\begin{equation}
\begin{aligned}
	\isdDot &= \frac{1}{\sigma\tauS}\left( \sigma\tauS\omdq\isq + \frac{\usd}{\Rs} - (1 - \sigma)\tauS\irdDot - \isd \right)\\
	\isqDot &= \frac{1}{\sigma\tauS}\left( -\sigma\tauS\omdq\isd + \frac{\usq}{\Rs} - (1 - \sigma)\tauS\ird - \isq \right)\\
	\irdDot &= \frac{\isd - \ird}{\tauR}\\
	\omdq &= p\omegaRMech + \omegaSl \quad \mathrm{with}\quad \omegaSl = \frac{\isq}{\tauR\ird} \\
    \Tmech &= \frac{3}{2}p\frac{\Lh}{1 + \sigmaR}\isq\ird \mathcomma
\end{aligned}
\end{equation}
where $\isd$ and $\isq$ denote the stator currents in dq components, and $\ird = \Lh\Psi_\mathrm{rd}$ is a magnetizing current composed of the rotor flux $\Psi_\mathrm{rd}$ and the magnetizing inductance $\Lh$. The stator and rotor time constants are given by $\tauS = \frac{\Ls}{\Rs}$ and $\tauR = \frac{\Lr}{\Rr}$, respectively, where $L$ and $R$ denote inductances and resistances. The rotor leakage coefficient is defined as the ratio of leakage inductance to magnetizing inductance, $\sigmaR = \frac{\LsigmaR}{\Lh}$. The stator leakage coefficient $\sigmaS$ follows analogously, leading to the total leakage coefficient $\sigma = 1 - \frac{1}{(1 + \sigmaS)(1 + \sigmaR)}$. The angular frequency of the dq reference frame $\omdq$ results from the pole pair number $p$, the mechanical speed $\omegaRMech$, and the angular frequency of the rotor currents $\omegaSl$. The input voltages in dq coordinates are $\usd$ and $\usq$, and the stator currents and the mechanical torque $\Tmech$ serve as output quantities.

For the second level of model fidelity, we include the skin effect in the rotor, which describes the increase in effective rotor resistance~\cite{binder_elektrische_2017}. To this end, the rotor time constant $\tauR$ is computed dynamically as a function of the rotor current frequency through
\begin{equation}
\begin{aligned}
    \SV &= h_\mathrm{c}\sqrt{\pi \mu_0\frac{b}{b_\mathrm{n}\rho}\omegaSl} \\
	\tauR &= \frac{\Lr}{\Rr}\Re\left\{ \SV\frac{\sinh (2\SV) + \sin (2\SV)}{\cosh (2\SV) - \cos (2\SV)} \right\} \mathdot
\end{aligned}
\end{equation}
The `reduced conductor height` $\SV$ is determined by the rotor conductor height $h_\mathrm{c}$, the vacuum permeability $\mu_0$, the rotor bar width $b$, the slot width $b_\mathrm{n}$, and the specific resistance of the conductor material $\rho$. The symbol $\Re\{\cdot\}$ denotes taking the real part of its argument

For the third level of model fidelity, we additionally account for saturation, which increases the stator leakage inductance at low speeds. To capture this, the stator leakage inductance as a function of speed was determined from measurements and incorporated into the model via a lookup table $\LsigmaS = f(\omegaRMech)$.

\subsubsection{Electrical Grid}
\label{subsec:grid-model}
The grid is modeled to reproduce the electrical boundary conditions required for testing the \gls{wea} in accordance with FGW TR3. As the relevant test cases include steady-state operation, voltage unbalance, harmonic distortion, and fault conditions, a single grid representation is insufficient. Therefore, we use four complementary grid models, each capturing a specific class of grid phenomena.

The `standard model` is based on a three-phase controlled voltage source with variable voltage magnitude, frequency, and phase angle. Voltage and frequency variations can be applied as step, ramp, or time-series signals, enabling the reproduction of grid-code-relevant operating conditions, including voltage deviations of up to $\pm 5\%$ and frequency variations around the nominal value. The grid frequency is determined using a PLL-based estimator.

Voltage unbalance is introduced by extending the model to generate positive- and negative-sequence voltage components via symmetrical component decomposition. The activation of unbalanced conditions is time-controlled, allowing the emulation of both transient and steady-state scenarios.

We represent harmonic distortion by superimposing harmonic components on the fundamental voltage, considering harmonics of configurable magnitude and phase up to the 50th order. Both positive- and negative-sequence harmonic components are included, enabling investigation of harmonic interactions and their impact on the \gls{wea} behavior.

For fault ride-through studies, a dedicated model is implemented in which the grid impedance is explicitly defined for two-phase and three-phase faults as a function of the applied voltage dip. The impedance values are selected based on the fault type and depth, thereby reproducing realistic fault conditions, including the associated voltage-magnitude and phase-angle behavior, consistent with field measurements and laboratory test benches \cite{jassmann_certbench_2021}.

\subsubsection{Load Application Model}
On the generator side, the load application is performed by the current-controlled test bench converter. The current controller employs an observer-based, model-predictive approach developed by dSPACE. To model this system as a transfer function, the step response to a current reference step was determined. The resulting current controller is represented as a combination of a first-order lag element with time constant $\Tcc = 2\pi \SI{80}{\kilo\hertz}$ and a multiplicative time delay of $\Td = \SI{2.4}{\micro\second}$.

For grid emulation, a superimposed voltage controller is required, since the test bench must regulate the grid voltage at the \gls{poc}. As the voltage controller operates at a very small sampling time of \SI{24}{\nano\second}, while the fundamental frequency of the emulated grid is \SI{50}{\hertz}, a separate PI controller is used for each phase.

\subsubsection{Model of the Device Under Test}
\begin{figure}[!t]
    \centering
	\includegraphics[width=\columnwidth]{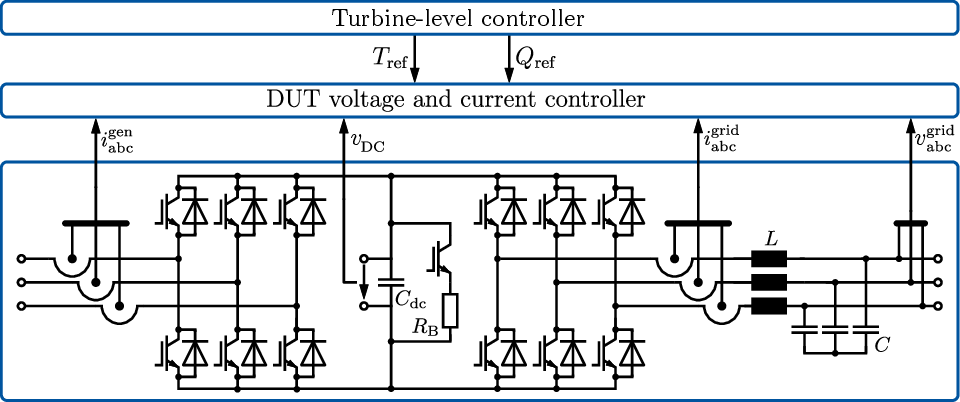}
	\caption{Schematic layout of the \gls{dut}. The break-chopper resistance $\rbrake$ is required to dissipate energy during fault ride-through test cases.}
	\label{fig:igbt-full-scale-converter}
\end{figure}

The \gls{dut} comprises the turbine-level controller and the full-scale power electronics converter with its associated control, as illustrated schematically in Fig.~\ref{fig:igbt-full-scale-converter}.

The primary function of the turbine-level controller is to regulate the rotor speed via the pitch angle, while the power extraction from the wind is determined through a torque setpoint for the generator using a precomputed lookup table. The turbine-level controller thus corresponds to a simplified yet industry-standard turbine control scheme~\cite{jonkman_definition_2009}. In addition, the static setpoint for reactive power is specified by the corresponding test case definition in the FGW TR3.

From a power electronics perspective, the \gls{dut} consists of two three-phase converters coupled through a shared DC link: The machine-side converter controls the excitation and torque of the generator, while the grid-side converter regulates the DC-link voltage and the reactive power at the \gls{poc}. The converter's harmonic emissions are attenuated by an LCL filter. Here, the grid-side inductance is realized through the leakage inductance of the nacelle transformer, which is reproduced by the real-time simulation of the grid.

The torque control of the machine-side converter is based on field-oriented control in a rotating reference frame aligned with the generator's rotor flux. The generator speed is provided through an encoder interface between the test bench and the \gls{dut} \cite{rw_de_doncker_advanced_2020}. The torque setpoint, supplied by the turbine-level controller, is converted into a corresponding current reference using the machine model equations.

The grid-side converter employs a voltage-oriented control strategy synchronized with the grid voltage. For grid synchronization, a \gls{ddsrf} is used, which synthesizes the positive- and negative-sequence components \cite{teodorescu_grid_2011}. Accordingly, the converter operates in grid-following mode and can regulate both active and reactive power independently, where the active power fed into the grid serves to regulate the DC-link voltage. The parameters of the full-scale converter are listed in Table \ref{tab:dut-table}.

\begin{table}[!t]
    \centering
    \caption{Electrical parameters of the \gls{dut}.}
    \label{tab:dut-table}
    \begin{tabular}{l|l|l} 
    \hline
    \textbf{Parameter} & \textbf{Symbol} & \textbf{Value}\\
    \hline
    Rated output current (rms) & $\irms$ & $1200 \, \mathrm{A}$\\
    Switching frequency & $\fsw$ & $2.5 \, \mathrm{kHz}$\\
    DC voltage & $\udc$ & $1100 \, \mathrm{V}$\\
    DC-link capacitance & $\cdc$ & $35 \, \mathrm{mF}$\\
    Brake chopper resistance & $\rbrake$ & $290 \, \mathrm{m\Omega}$\\
    \hline
    \end{tabular}
\end{table}

\subsubsection{Geographic Distribution Model}
To model the geographic distribution, we first consider the machine side. The geographic distribution on this side arises from outsourcing the aeroelastic simulation, which connects to the remaining test bench via the internet. The interface is therefore characterized by the prevailing time delay. Since the internet time delay is non-deterministic, we model it as a random variable with a specified mean and standard deviation.

On the grid side, the interface was realized through a direct fiber-optic connection between the two sites. This is necessary because the grid dynamics tolerate only time delays on the order of a few hundred microseconds, rendering transmission over the internet infeasible. Since the fiber-optic connection establishes a direct FPGA-to-FPGA communication link, only minimal deviations of the time delay from its mean value occur. Accordingly, we model the time delay on this side as a fixed value.

\section{Test Bench Setup}
\label{sec:test-bench-setup}
\subsection{Overview}
\begin{figure*}[!t]
	\centering
	\includegraphics[width=\textwidth]{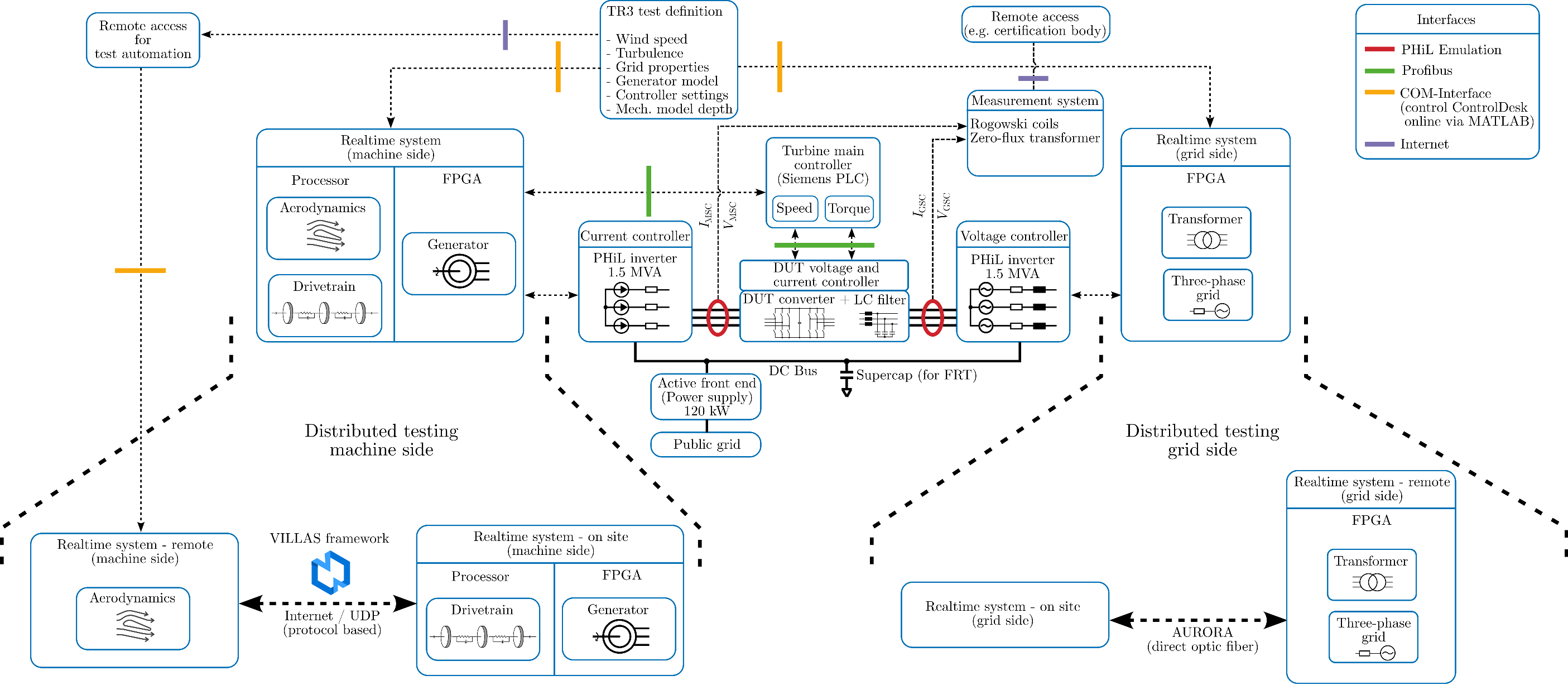}
	\caption{Overview of the complete setup of the proposed test bench.}
	\label{fig:testbench-layout-complete}
\end{figure*}

The hardware layout of the test bench is shown in Fig. \ref{fig:testbench-layout-complete}. The machine-side and grid-side emulators each consist of 48 dSPACE DS5386 modules, enabling tests at up to $1.5\,\mathrm{MVA}$. Integrating the \gls{dut} between the two emulators allows for a circular power flow within the system, so that only electrical losses need to be compensated by the \gls{afe}, i.e., the power supply unit of the test bench. This permits the power supply to be dimensioned much smaller than the \gls{dut} rated power. The parameters of the test bench are listed in Table~\ref{tab:TB-table}.

\begin{table}[!ht]
    \caption{Parameters of the test bench.}
    \label{tab:TB-table}
    \centering
    \begin{tabular}{l|l|l} 
    \hline
    \textbf{Parameter} & \textbf{Symbol} & \textbf{Value}\\
    \hline
    Rated power & $S_\mathrm{N}$ & $1.5 \, \mathrm{MVA}$ \\
    Supply power (without auxiliaries) & $P_\mathrm{supply}$ & $120 \, \mathrm{kW}$ \\
    Number of modules & $N_\mathrm{DS5386}$ & $96$ \\
    Rated output current per module (AC rms) & $\irms$ & $75 \, \mathrm{A}$ \\
    DC voltage & $\utb$ & $1500 \, \mathrm{V}$ \\
    \hline
    \end{tabular}
\end{table}

\subsection{FRT Testing with Supercapacitors}
\gls{frt} tests pose a particular challenge to the circular power flow concept. This arises because \gls{frt} tests require voltage dips down to $30 \, \%$ of the nominal voltage, with optional dips down to 0~p.u. Grid connection rules require that the \gls{dut} detects a voltage dip, maintains operation for a minimum duration, and provides reactive current to support the grid, where the exact duration and amount depend on the depth of the voltage dip. Moreover, reactive current injection during the \gls{frt} event takes priority over active power feed-in. The latter must therefore be reduced if necessary to avoid exceeding the maximum output current. As the drivetrain has high inertia, the generator power cannot be reduced with the same dynamics as the reduced active power feed-in to the grid, leading to an increase in the DC-link voltage. Therefore, due to the generator input power and the lack of active-power injection on the grid side, a large portion of the turbine power is dissipated in the brake chopper as heat and must therefore be supplied by the power supply. However, the system's power supply is not dimensioned for this test case, so its rated power must either be increased at considerable additional cost, or the DC-link capacitance of the test bench must be enlarged to provide sufficient energy to sustain the \gls{frt} test.

To ensure that sufficient energy is available in the test bench DC link for \gls{frt} tests, the DC-link capacitance was realized using a stack of modular supercapacitors, connecting a total of ten modules, each with a rated capacitance of \SI{66}{\farad} and a rated voltage of \SI{162}{\volt}, in series. The resulting total DC-link capacitance of $6.6 \mathrm{F}$ enables FRT testing despite a compact, cost-efficient power supply.

\subsection{Real-Time System}
The real-time system of the test bench comprises a dSPACE SCALEXIO system, featuring a Processing Unit (PU) with three usable cores and two DS6601 FPGAs. The mechanical models, the test bench control and monitoring logic, and the grid model run on the PU. Since the generator models require sampling times of at most \SI{1}{\micro\second}, they were implemented on the FPGA using Xilinx System Generator and validated in advance through step-response tests to ensure equivalent behavior to the reference model. The current controller is integrated on the DS5386 load application modules, and the voltage controllers for grid emulation run on the second FPGA. The development and implementation of the overall system are carried out in Simulink and compiled into a real-time executable using the dSPACE toolchain, comprising ConfigurationDesk and ControlDesk, before deployment to the real-time system.

\subsection{Geographic Distribution}
\begin{figure}[!t]
	\centering
	\includegraphics[width=\columnwidth]{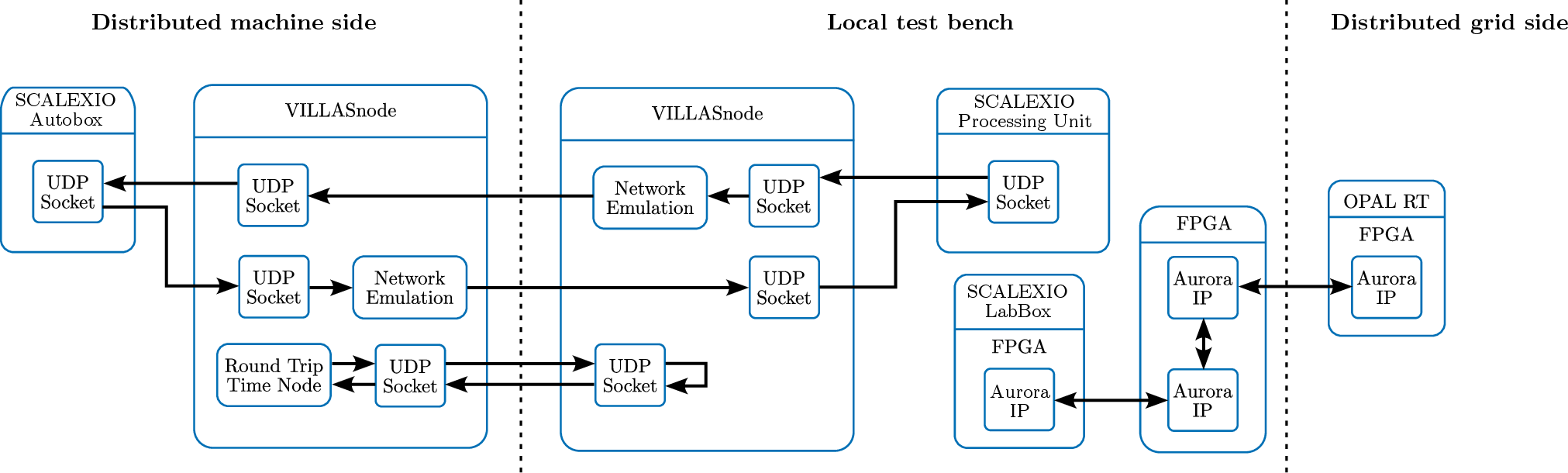}
	\caption{Communication setup between the three sites.}
	\label{fig:comm_setup}
\end{figure}

The communication setup between the three sites is illustrated in Fig.~\ref{fig:comm_setup}. As described above, the geographic distribution on the machine side is realized via the internet. At the distributed machine side, the aerodynamic model runs on a dSPACE SCALEXIO Autobox.
The interface to the test bench is implemented using VILLASnode~\cite{vogel_villasnode_2025}, which essentially functions as a protocol translator and runs on a Raspberry Pi at each site.
Incoming messages from the real-time systems are forwarded via UDP to the other site, thereby creating a virtual subnet that interconnects the two sites.

Due to the time-varying nature of the communication delay, older UDP messages may arrive before newer ones. This poses a problem for real-time simulation, since dynamic systems are modeled at both ends of the connection, and the temporal order of input signals matters. To ensure the correct temporal sequence, a unique message ID is generated at the test bench and appended to each message. The distributed machine side attaches the same ID to its response. Since the temporal order of the IDs is known at the local test bench, this mechanism ensures that no outdated messages are processed. Additionally, the time offset between sending and receiving is used to measure the current latency in the distributed system, ensuring that tests conducted under elevated latency are not inadvertently deemed valid.

On the grid side, communication is implemented using the AURORA protocol~\cite{amd_aurora_2025}, employing the 8b10b variant.
The grid simulator on the distributed grid side is connected to the local test-bench site via a single-mode fiber-optic link, yielding a round-trip latency of \SI{160}{\micro\second}, which is well below an analytically determined limit of \SI{350}{\micro\second}.
At the local test bench, an FPGA performs the conversion between single-mode and multi-mode fiber, since the dSPACE LabBox supports only multi-mode connections.

\section{Preliminary Results and Discussion}
\label{sec:preliminary-results-and-discussion}
\subsection{Machine-Side Geographic Distribution: Latency Sensitivity Analysis}
\label{subsec:machine-side-latency-sensitivity-analysis}
A latency sensitivity analysis of the machine-side geographically distributed system was carried out experimentally using the assembled test bench to determine the influence of communication delay. We conducted the study on signal level, i.e., in closed-loop simulation with a simulated \gls{dut}, but with the fully operable geographically distributed system.

\begin{figure}[!t]
    \centering
    \includegraphics[width=\columnwidth]{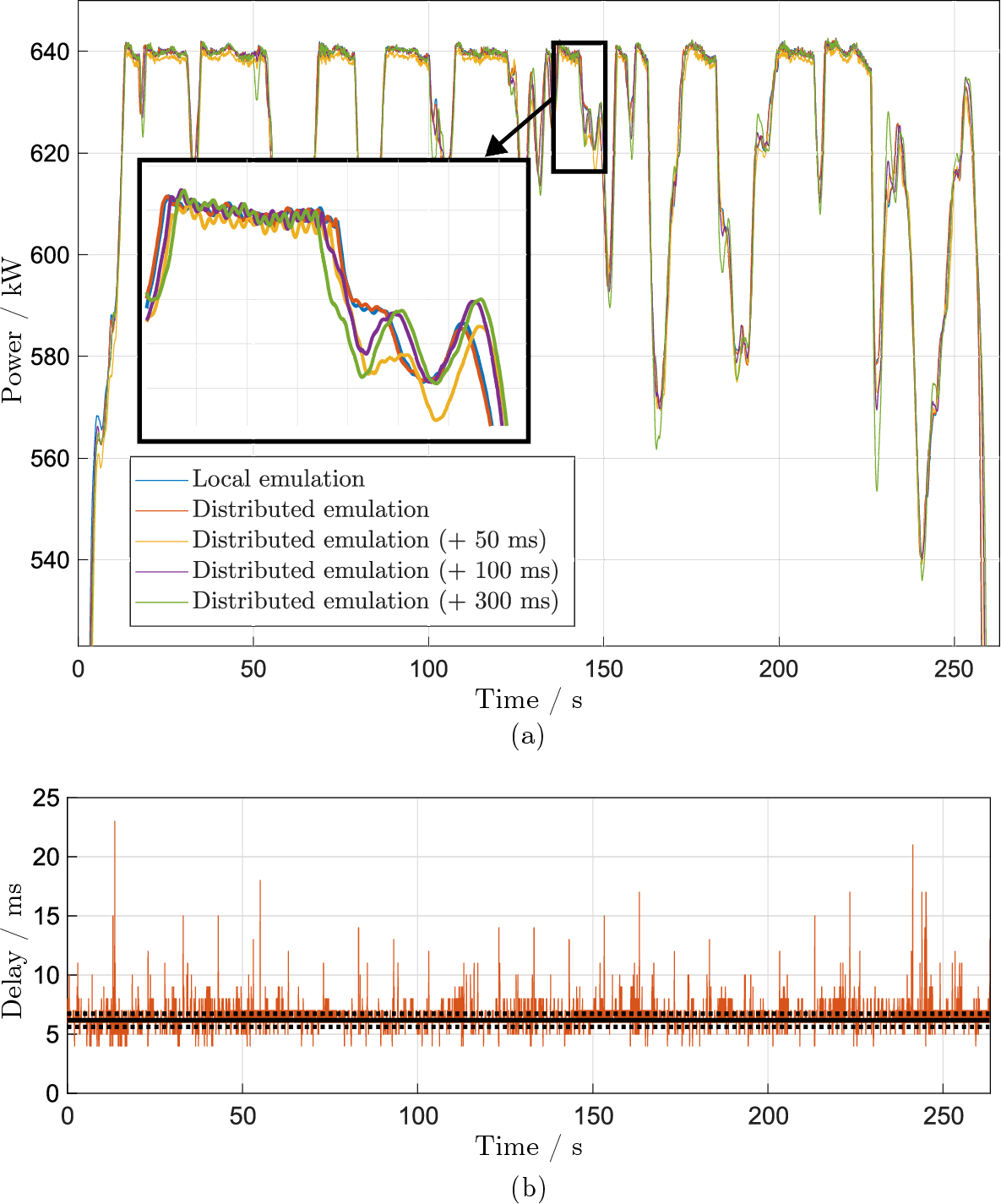}
    \caption{Sensitivity of the machine-side system to time delay in the distributed system (a) and measured time delay during nominal operation (b). In (b), the mean and standard deviation are indicated by solid and dashed lines, respectively.}
    \label{fig:machine-side-latency-sensitivity}
\end{figure}

The results of this investigation are shown in Fig.~\ref{fig:machine-side-latency-sensitivity}. Since the mean latency during operation lies below \SI{10}{\milli\second}, additional latency was introduced in software for the analysis. Exemplified by the generated mechanical power, the results show that under nominal operation, i.e., without artificially added latency, the power is nearly identical to the purely local emulation, while pronounced deviations become apparent at higher latency values.

A quantitative evaluation yields the relative total spectral distortion $\DistSpec$, which is defined by 
\begin{equation}
    \DistSpec = \sum_{k=0}^{N} \left| \frac{A(f_k) - A_\mathrm{ref}(f_k)}{A_\mathrm{ref}(f_k)} \right| \cdot 100\% \mathcomma
\end{equation}
where $A(f_k)$ denotes the single-sided power spectrum of the mechanical power, computed via \gls{fft}, and $f_k \, \mathrm{with} \, k \in [0,N]$ are the individual frequency bins~\cite{duhamel_fast_1990}. The reference spectrum $A_\mathrm{ref}(f_k)$ was obtained by purely local simulation. The results, reported in Table~\ref{tab:relative-total-spectral-distortion}, show that the total spectral distortion is lowest for the case without added time delay, and already exceeds 10\,\% at \SI{50}{\milli\second}. As the closed-loop system involves nonlinear parts, $\DistSpec$ does not grow linearly with the added delay.
\begin{table}[!ht]
    \centering
    \caption{Relative total spectral distortion in the distributed system, considering frequencies up to \SI{500}{\hertz}.}
    \label{tab:relative-total-spectral-distortion}
    \begin{tabular}{c|c}
    \hline
    Added delay (ms) & $\DistSpec$ \\
    \hline
     0 & 4.22 \% \\
     50 & 14.55 \% \\
     100 & 13.84 \% \\
     300 & 19.94 \% \\
     \hline
    \end{tabular}
\end{table}

Overall, we conclude that the nominal geographically distributed emulation, i.e., with a round-trip time delay of \SI{6}{\milli\second} and no additional time delay, achieves good emulation quality due to the small spectral distortion. For tests with the geographically distributed system, the time delay should remain below \SI{10}{\milli\second} in both the mean and the standard deviation.

\subsection{Quality of the Grid Emulation}
\label{subsec:grid-emulation-quality}
\begin{figure}[!t]
    \centering
    \includegraphics[width=\columnwidth]{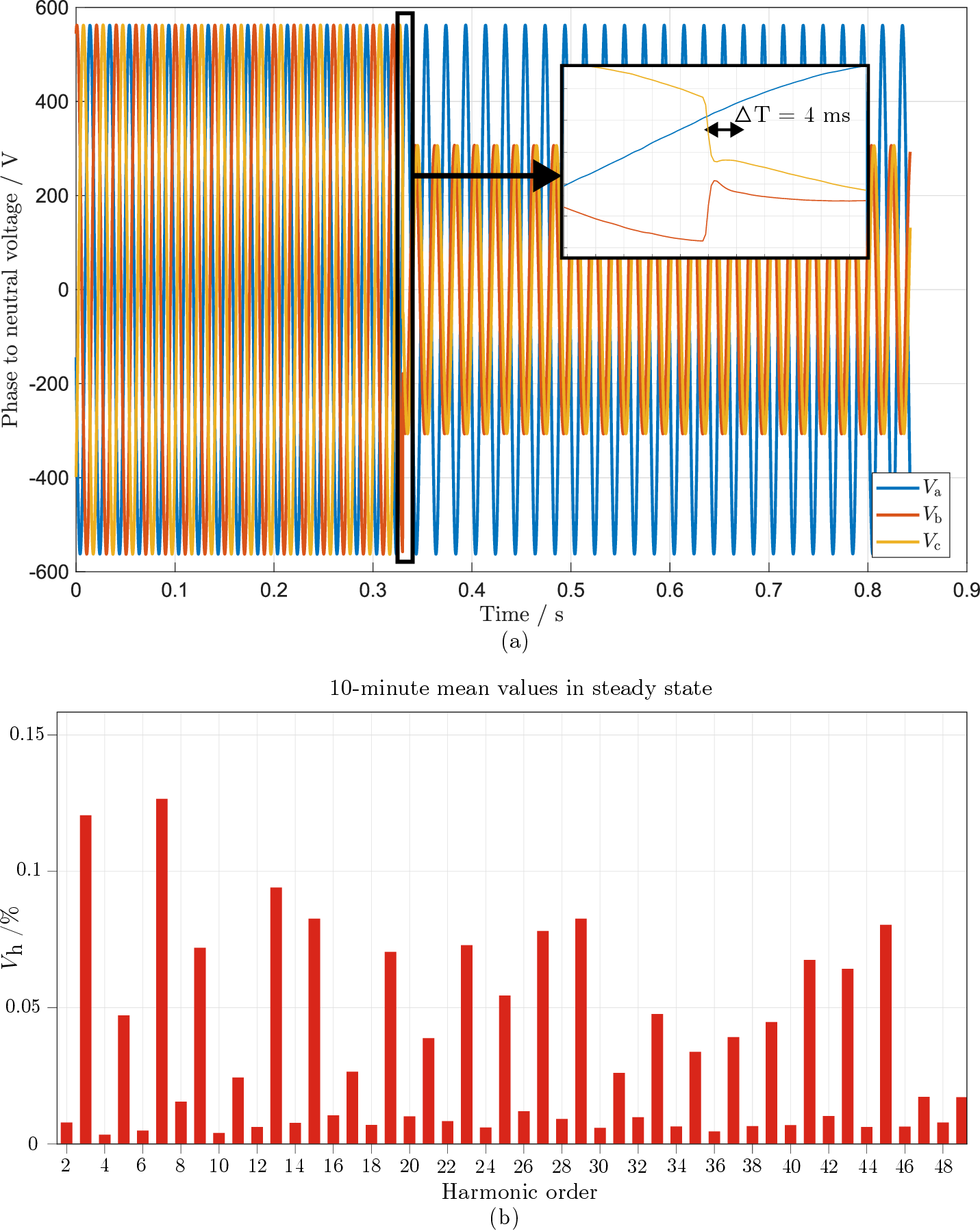}
    \caption{Unsymmetrical voltage dip to $0.5\,\mathrm{p.u.}$ reproduced
    by the grid emulator during no-load fault ride-through validation (a) and harmonic components of the emulated voltage in steady state (b).}
    \label{fig:frt-symmetrical-harmonics}
\end{figure}

For the experimental no-load validation of the grid models, we exemplify the grid emulator's fault ride-through capability. No-load refers to test-bench operation without the \gls{dut}, i.e., voltages were measured at the open-circuit terminals of the grid emulator interface with no current flowing. Results for an asymmetrical two-phase voltage dip to $0.5\,\mathrm{p.u.}$ are shown in Fig.~\ref{fig:frt-symmetrical-harmonics}~(a). The voltage dip is applied through the impedance variation of the FRT grid model, and the resulting per-phase voltage magnitudes and phase angles are determined according to the Bollen classification of unbalanced voltage dips~\cite{bollen_understanding_nodate}, ensuring compatibility with test procedures based on physical fault reproduction. The results show the pure voltage-control performance of the grid emulator, i.e., the physical behavior of a circuit breaker is not emulated here, but will be added in the future. As no current was flowing, the capability of grid impedance emulation is not evident in the presented results, although it remains present in the real-time system.

The harmonic content of the emulated grid voltage under steady-state conditions is shown in Fig.~\ref{fig:frt-symmetrical-harmonics}~(b) for Phase~a ($V_\mathrm{a}$), presenting the 10-minute mean value of each harmonic order up to the 50th, normalized to the 
nominal voltage. As higher-order harmonics remain well below $0.2\,\%$, this demonstrates the excellent voltage quality achieved by the grid emulator.

Additionally, the measurements were evaluated against the requirements for voltage quality assessment in FGW TR3, with the results reported in Table~\ref{tab:grid-quality-measurements}. All specified requirements are fulfilled.

\begin{table}[!ht]
    \centering
    \caption{Results from the no-load grid emulation quality
    measurement, computed as 10-minute mean values. HF refers to high-frequency components in the range from 2-9~kHz.}
    \label{tab:grid-quality-measurements}
    \begin{tabular}{>{\raggedright\arraybackslash}p{0.33\columnwidth}|c|c}
    \hline
    \textbf{Quantity} & \textbf{Requirement} & \textbf{Result} \\
    \hline
    THDS (voltage)
        & $\leq 5\,\%$
        & $0.33\,\%$ \\
    THDS (voltage, HF)
        & $\leq 3\,\%$
        & $0.17\,\%$ \\
    Voltage deviation from 1~p.u.
        & max. $\pm 10\,\%$
        & mean: 0.997~p.u. \\
    Frequency deviation from 50~Hz
        & max. $\pm 1\,\%$
        & mean: 50~Hz\\
    ROCOF
        & max. $\pm 0.2\,\%\,/\,0.2\,\mathrm{s}$
        & $<$ 0.01 \% / 0.2\, s \\
    Voltage asymmetry
        & $\leq 2\,\%$
        & $0.03\,\%$ \\
    \hline
    \end{tabular}
\end{table}

\section{Conclusion and Outlook}
\label{sec:conclusion-and-outlook}
This paper presented the concept, design, modeling, and construction of a megawatt-class PHiL component test bench equipped with geographically distributed simulation for grid-compliance testing of wind turbines. The test bench replaces the physical generator and the electrical grid with power electronics-based real-time emulation, enabling norm-based determination of the electrical properties of wind turbines on a PHiL component test bench. A two-stage emulation concept was introduced, combining virtualization and emulation of non-physical components with geographically distributed real-time simulation of the aerodynamics and the grid model.

Preliminary experimental results are promising. As the machine- and grid-side emulation quality is adequate for high-precision tests, this presents the opportunity of carrying out sensitivity analyses with the full system, ranging from generator model depth to distributed system latency. As their influence on certification-relevant tests is still unclear, we expect that upcoming closed-loop tests will help further establish component testing within the industry and possibly improve technical specifications and standards regarding required or allowed boundary conditions for wind turbine testing.

\ifCLASSOPTIONcaptionsoff
  \newpage
\fi

\bibliographystyle{IEEEtran}
\bibliography{bibliography/references}

@article{neshati_grid_nodate,
	title = {Grid connection testing of wind energy converters with medium voltage generator- and grid emulation on a multi-megawatt power-hardware-in-the-loop test rig},
	url = {https://digital-library.theiet.org/doi/10.1049/icp.2023.2773},
	language = {en},
	journal = {IET Conference Proceedings},
    year = {2023},
	author = {Neshati, Mosen},
}

@misc{european_commission_commission_2016,
	title = {Commission {Regulation} ({EU}) 2016/631 of 14 {April} 2016 establishing a network code on requirements for grid connection of generators},
	url = {http://data.europa.eu/eli/reg/2016/631/oj},
	author = {{European Commission}},
	month = apr,
	year = {2016},
}

@misc{noauthor_verordnung_2017,
	title = {Verordnung zum {Nachweis} von elektrotechnischen {Eigenschaften} von {Energieanlagen} ({Elektrotechnische}-{Eigenschaften}-{Nachweis}-{Verordnung} - {NELEV})},
	url = {https://www.gesetze-im-internet.de/nelev/BJNR165100017.html},
	month = jun,
	year = {2017},
	pages = {1651},
}

@misc{vde_verband_der_elektrotechnik_elektronik_informationstechnik_ev_technical_2023,
	title = {Technical {Connection} {Rules} for {Medium}-{Voltage} ({Technische} {Anschlussregeln} {Mittelspannung})},
	url = {https://www.vde.com/de/fnn/themen/tar/tar-mittelspannung-vde-ar-n-4110},
	author = {{VDE Verband der Elektrotechnik Elektronik Informationstechnik e.V.}},
	month = sep,
	year = {2023},
}

@book{binder_elektrische_2017,
	address = {Berlin, Heidelberg},
	title = {Elektrische {Maschinen} und {Antriebe}: {Grundlagen}, {Betriebsverhalten}},
	copyright = {https://www.springer.com/tdm},
	shorttitle = {Elektrische {Maschinen} und {Antriebe}},
	url = {https://link.springer.com/10.1007/978-3-662-53241-6},
	language = {de},
	urldate = {2026-04-06},
	publisher = {Springer},
	author = {Binder, Andreas},
	year = {2017},
	doi = {10.1007/978-3-662-53241-6},
}

@book{gerling_electrical_2015,
	address = {Berlin, Heidelberg},
	series = {Mathematical {Engineering}},
	title = {Electrical {Machines}: {Mathematical} {Fundamentals} of {Machine} {Topologies}},
	volume = {4},
	copyright = {https://www.springernature.com/gp/researchers/text-and-data-mining},
	shorttitle = {Electrical {Machines}},
	url = {https://link.springer.com/10.1007/978-3-642-17584-8},
	language = {en},
	urldate = {2026-04-06},
	publisher = {Springer},
	author = {Gerling, Dieter},
	year = {2015},
	doi = {10.1007/978-3-642-17584-8},
}

@techreport{fgw_ev_technical_2022,
	address = {Berlin},
	title = {Technical {Guidelines} for {Power} {Generating} {Units} and {Systems}, {Part} 3: {Determination} of the {Electrical} {Characteristics} of {Power} {Generating} {Units} and {Systems} in {Medium}-, {High}- and {Extra}-{High} {Voltage} {Grids}},
	number = {TR 3 Rev. 26},
	institution = {{FGW e.V. – Fördergesellschaft Windenergie und andere Dezentrale Energien}},
	author = {{FGW e.V.}},
	month = apr,
	year = {2022},
}

@techreport{fgw_ev_technical_2019,
	address = {Berlin},
	title = {Technical {Guidelines} for {Power} {Generating} {Units} and {Systems}, {Part} 8: {Certification} of the {Electrical} {Characteristics} of {Power} {Generating} {Units} and {Plants} in {Medium}-, {High}- and {Extra}-{High} {Voltage} {Grids}},
	number = {TR 8 Rev. 9},
	institution = {{FGW e.V. – Fördergesellschaft Windenergie und andere Dezentrale Energien}},
	author = {{FGW e.V.}},
	month = feb,
	year = {2019},
}

@techreport{international_electrotechnical_commission_wind_2025,
	address = {Geneva},
	title = {Wind energy generation systems – {Part} 21-4: {Measurement} and assessment of electrical characteristics – {Wind} turbine components and subsystems},
	number = {IEC TS 61400-21-4:2025},
	institution = {{International Electrotechnical Commission (IEC)}},
	author = {{International Electrotechnical Commission}},
	year = {2025},
}

@inproceedings{ren_improve_2007,
	address = {Edmonton, AB, Canada},
	title = {Improve the {Stability} and the {Accuracy} of {Power} {Hardware}-in-the-{Loop} {Simulation} by {Selecting} {Appropriate} {Interface} {Algorithms}},
	doi = {10.1109/ICPS.2007.4292112},
	booktitle = {{IEEE}/{IAS} {Industrial} \& {Commercial} {Power} {Systems} {Technical} {Conference}},
	author = {Ren, W. and Steurer, M. and Baldwin, T. L.},
	year = {2007},
	pages = {1--7},
}

@article{marnett_real-time_2014,
	title = {Real-time simulation of aeroelastic rotor loads for horizontal axis wind turbines},
	volume = {524},
	issn = {1742-6596},
	url = {https://iopscience.iop.org/article/10.1088/1742-6596/524/1/012071},
	doi = {10.1088/1742-6596/524/1/012071},
	number = {1},
	journal = {Journal of Physics: Conference Series},
	author = {Marnett, M. and Wellenberg, S. and Schröder, W.},
	month = jan,
	year = {2014},
	note = {Publisher: IOP Publishing},
	pages = {012071},
}

@phdthesis{uwe_jassmann_hardware---loop_2018,
	address = {Aachen},
	title = {Hardware-in-the-{Loop} {Wind} {Turbine} {System} {Test} {Benches} and their {Usage} for {Controller} {Validation}},
	url = {http://publications.rwth-aachen.de/record/750754/files/750754.pdf},
	author = {Uwe Jassmann},
	collaborator = {RWTH Aachen University},
	month = jan,
	year = {2018},
}

@article{jassmann_certbench_2021,
	title = {{CertBench}: conclusions from the comparison of certification results derived on system test benches and in the field},
	volume = {85},
	issn = {0015-7899},
	url = {https://publications.rwth-aachen.de/record/817412/files/817412.pdf},
	doi = {10.1007/s10010-021-00470-1},
	number = {2},
	journal = {Forschung im Ingenieurwesen},
	author = {Jassmann, Uwe and Frehn, Anica and Röttgers, Heiko and Santjer, Fritz and Mehler, Christian and Beißel, Thomas and Kaven, Lennard and den Hoff, Daniel and Frühmann, Richard and Azarian, Soroush and Neshati, Mohsen and Quistorf, Gesa and Zuga, Adam and Brennecke, Martin and Jersch, Torben and Schelenz, Ralf},
	month = jan,
	year = {2021},
	note = {Num Pages: 19},
	pages = {353--371},
}

@article{orourke_geometric_2019,
	title = {A {Geometric} {Interpretation} of {Reference} {Frames} and {Transformations}: dq0, {Clarke}, and {Park}},
	volume = {34},
	issn = {0885-8969},
	doi = {10.1109/TEC.2019.2941175},
	number = {4},
	journal = {IEEE Transactions on Energy Conversion},
	author = {O'Rourke, Colm J. and Qasim, Mohammad M. and Overlin, Matthew R. and Kirtley, James L.},
	month = jan,
	year = {2019},
	pages = {2070--2083},
}

@book{rw_de_doncker_advanced_2020,
	title = {Advanced {Electrical} {Drives}: {Analysis}, {Modeling}, {Control}},
	publisher = {Springer International Publishing},
	author = {R.W. de Doncker and Duco W.J. Pulle and Andre Veltman},
	month = jan,
	year = {2020},
}

@techreport{jonkman_definition_2009,
	address = {Golden, Colorado},
	title = {Definition of a 5-{MW} {Reference} {Wind} {Turbine} for {Offshore} {System} {Development}},
	url = {https://www.nrel.gov/docs/fy09osti/38060.pdf},
	number = {NREL/TP-500-38060},
	institution = {NREL},
	author = {Jonkman, J. and Butterfield, S. and Musial, W. and Scott, G.},
	month = feb,
	year = {2009},
	note = {Num Pages: 75},
}

@book{teodorescu_grid_2011,
	title = {Grid {Converters} for {Photovoltaic} and {Wind} {Power} {Systems}},
	isbn = {978-0-470-66705-7},
	language = {en},
	urldate = {2026-04-06},
	publisher = {John Wiley \& Sons, Ltd},
	author = {Teodorescu, Remus and Liserre, Marco and Rodríguez, Pedro},
	year = {2011},
	doi = {10.1002/9780470667057.fmatter},
	note = {\_eprint: https://onlinelibrary.wiley.com/doi/pdf/10.1002/9780470667057.fmatter},
}

@article{duhamel_fast_1990,
	title = {Fast fourier transforms: {A} tutorial review and a state of the art},
	volume = {19},
	issn = {0165-1684},
	shorttitle = {Fast fourier transforms},
	url = {https://www.sciencedirect.com/science/article/pii/016516849090158U},
	doi = {10.1016/0165-1684(90)90158-U},
	number = {4},
	urldate = {2026-04-06},
	journal = {Signal Processing},
	author = {Duhamel, P. and Vetterli, M.},
	month = apr,
	year = {1990},
	pages = {259--299},
}

@article{vogel_villasnode_2025,
	title = {{VILLASnode}: {An} {Open}-{Source} {Real}-time {Multi}-protocol {Gateway}},
	volume = {10},
	doi = {10.21105/joss.08401},
	number = {112},
	journal = {Journal of Open Source Software},
	author = {Vogel, Steffen and Eiling, Niklas and Pitz, Manuel and Bach, Alexandra and Stevic, Marija and Monti, Antonello},
	year = {2025},
	pages = {8401},
}

@misc{amd_aurora_2025,
	title = {Aurora {8B}/{10B} {LogiCORE} {IP} {Product} {Guide} ({PG046})},
	author = {AMD},
	year = {2025},
}

@book{bollen_understanding_nodate,
	title = {Understanding {Power} {Quality} {Problems}},
	url = {https://www.wiley-vch.de/de/fachgebiete/ingenieurwesen/understanding-power-quality-problems-978-0-7803-4713-7},
	language = {de-de},
	urldate = {2026-04-06},
	publisher = {IEEE Press Series on Power Engineering},
	author = {Bollen, Math H.J.},
}

\raggedbottom
\begin{IEEEbiographynophoto}{Simon Moritz}
received the M. Sc. degree in Electrical Power Engineering from RWTH Aachen University in 2023.
Since 2023, he has been a research associate at the Institute of Automatic Control, RWTH Aachen University, Germany, working towards a Ph.D. in Control Engineering. His research interests include Power Hardware-in-the-Loop simulation of wind energy systems, control system design, and robust control of drive systems.
\end{IEEEbiographynophoto}

\begin{IEEEbiographynophoto}{Yoga Viknesh Kannan}
received the M.Sc. degree in Electrical Power Engineering from RWTH Aachen University, Aachen, Germany, in 2022. He is currently a Research Associate with the E.ON Energy Research Center, Institute for Automation of Complex Power Systems, RWTH Aachen University. His research interests include modeling, control, and stability analysis of power electronics-driven power systems and Power Hardware-in-the-Loop studies.
\end{IEEEbiographynophoto}

\begin{IEEEbiographynophoto}{Florian Andresen}
earned his M.Sc. degree in electrical power engineering from RWTH Aachen University, Aachen, Germany, in 2022. He is working toward his Ph.D. degree at the E.ON Energy Research Center, Institute for Power Generation and Storage Systems, RWTH Aachen University, Germany.  His research interests include power electronics, PHiL testing, and wind energy systems.
\end{IEEEbiographynophoto}

\begin{IEEEbiographynophoto}{Laura Fuentes Grau}
received her M.Sc in computer engineering in 2023 from RWTH Aachen University. She is currently employed as a research associate with the Institute for Automation of Complex Power Systems, RWTH Aachen University, where she is also working towards her Ph.D. degree. Her research interests include digital hardware design, high-performance and real-time communication, as well as special operating systems.
\end{IEEEbiographynophoto}

\begin{IEEEbiographynophoto}{Manuel Fischer}
studied electrical engineering at the University of Stuttgart, where he received his Bachelor of Science degree in 2012 and his Master of Science degree in 2015. Following, he worked six years as a research associate at the Institute for Power Electronics and Electrical Drives at the University of Stuttgart (ILEA) and received the PhD degree in 2023. His PhD topic was the precise emulation of three-phase machines. Since 2021, he has been employed at dSPACE SE \& Co.\ KG as a product manager for Power HIL systems.
\end{IEEEbiographynophoto}

\begin{IEEEbiographynophoto}{Johannes-Martin Hördler}
graduated from University of Applied Sciences South Westfalia, Soest, Germany, in 2009 with a Dipl.-Ing. (FH) in Electrical Engineering and M. Sc. in Systems Engineering and Engineering Management with distinction in 2010 in cooperation with the University of Bolton, UK (dual-award). Since then, he worked at dSPACE SE \& Co.\ KG as application engineer, project manager and today as manager for project engineering for electromobility and electric drive applications. His focus is on Power HIL systems.
\end{IEEEbiographynophoto}

\begin{IEEEbiographynophoto}{Sabri Özer}
was born in 1993 in Silifke, Turkey. He studied Electrical Engineering at Karlsruhe Institute of Technology (KIT), from which he received his B.Sc. and M.Sc. Degrees in 2016 and 2020, respectively. Since 2023, he has been working at dSPACE SE \& Co.\ KG as a project engineer focusing on Power HIL systems.
\end{IEEEbiographynophoto}

\begin{IEEEbiographynophoto}{Martin Brennecke}
received his Dipl.-Ing. degree in Electrical Engineering from RWTH Aachen University in 2008. Since 2009, he has been with the FGH Research Association for Power Systems and Power Economics, where he has held various roles within the certification body and test laboratory. He began his career conducting LVRT tests with FGH’s mobile test laboratories and later developed FGH’s analytical software for assessing the FGW-TR4 compliance of simulation models, a key requirement for the certification of power generating units. Since 2013, he has served as an expert at the FGH Certification Office, carrying out assessments for product and unit certification. In this role, he has also contributed to the development of future DER model validation processes and simulation approaches for generation clusters such as wind farms. In 2016, he became Deputy Head of FGH’s Certification Body and Test Laboratory. Over the course of his career, he has actively participated in numerous working groups and standardization committees, including IEC TC88 MT 21, IEC TC88 WG 27, DKE 261.0.1, as well as FGW-TR8, FGW-TR4, and FGW-TR3.
\end{IEEEbiographynophoto}

\begin{IEEEbiographynophoto}{Philipp Kersting}
was born in 1997 in Hagen, Germany. He received his B. Sc. in Electrical Engineering, Information Technology and Computer Engineering in 2020 and an M. Sc. in Electrical Power Engineering in 2023 at RWTH Aachen University.  Following his graduation, he worked as research assistant at the Institute for Electrical Machines (IEM). Since 2025, he is employed at TIGRIS Elektronik GmbH as development engineer with a focus on power electronics.
\end{IEEEbiographynophoto}

\begin{IEEEbiographynophoto}{Andreas Thul}
received the Dipl.-Ing. degree in electrical engineering from RWTH Aachen University in 2013 and the Dr.-Ing. degree from the Faculty of Electrical Engineering and Information Technology at RWTH Aachen University in August 2021. He has been working as a research associate with the Institute of Electrical Machines at RWTH Aachen University since October 2013. From 2017 to 2021 he was the leader of the research group for mechatronics and drives at the IEM. Since November 2021 he is Chief Engineer at the IEM. His research interests include control and modelling of electrical machines and measurement techniques for the characterization of soft magnetic materials.
\end{IEEEbiographynophoto}

\begin{IEEEbiographynophoto}{Simon Steentjes}
received the Diploma and Dr.-Ing. degrees in electrical engineering and information technology from RWTH Aachen University, Aachen, Germany. From 2011 to 2018, he was a Research Associate and Group Leader with the Institute of Electrical Machines (IEM) at RWTH Aachen University. After completing his doctorate, he worked as a subject matter expert for the design of electric traction machines for xEV applications at Audi AG. He subsequently held leadership positions at the Hilti Group, where he was responsible for the multiphysical development and control of advanced electric drive systems. Since October 2024, he has been Full Professor of Electrical Machines and Drives and Director of the Institute of Electrical Machines (IEM) at RWTH Aachen University. He is the author or coauthor of more than 170 technical publications. His research focuses on the model-based analysis, design, and control of electrical drive systems, spanning multiphysical modeling, converter-machine interactions, and advanced control and estimation methods for energy conversion systems.
\end{IEEEbiographynophoto}

\begin{IEEEbiographynophoto}{Ferdinanda Ponci}
(SM’08) received her Ph.D. degree in electrical engineering from the Politecnico di Milano, Italy, in 2002. She joined the Department of Electrical Engineering, University of South Carolina, Columbia, SC, USA, as an Assistant Professor in 2003 and was tenured and promoted in 2008. In 2009, she joined the Institute for Automation of Complex Power Systems, E.ON Research Center, RWTH Aachen University, Aachen, Germany, where she is currently a Professor for Monitoring and Distributed Control for Power Systems. Her research interests include advanced measurement, monitoring and automation of active distribution systems. Prof. Ponci is a Senior Member of IEEE, an elected member of the Administration Committee of the IEEE Instrumentation and Measurement Society and the Liaison with IEEE Women in Engineering.
\end{IEEEbiographynophoto}

\begin{IEEEbiographynophoto}{Antonello Monti}
received his M.Sc degree (summa cum laude) and his PhD in Electrical Engineering from Politecnico di Milano, Italy, in 1989 and 1994, respectively. He started his career in Ansaldo Industria and then moved in 1995 to Politecnico di Milano as Assistant Professor.  In 2000, he joined the Department of Electrical Engineering of the University of South Carolina (USA) as Associate and then Full Professor.  Since 2008, he is the director of the Institute for Automation of Complex Power Systems within the E.ON Energy Research Center at RWTH Aachen University. Since 2019, he holds a double appointment with Fraunhofer FIT where he is developing the new Center for Digital Energy in Aachen. Dr. Monti is author or co-author of more than 400 peer-reviewed papers published in international Journals and in the proceedings of International conferences. He is a Senior Member of IEEE, , Associate Editor of IEEE Electrification Magazine, Member of the Editorial Board of the Elsevier Journal SEGAN and member of the founding board of the Springer Journal “Energy Informatics”. Dr. Monti was the recipient of the 2017 IEEE Innovation in Societal Infrastructure Award.
\end{IEEEbiographynophoto}

\begin{IEEEbiographynophoto}{Rik W. De Doncker}
received the M.Sc. degree in electromechanical power engineering and the Ph.D. degree in electrical power engineering (with the highest distinction) from the Katholieke Universiteit Leuven, Belgium, in 1981 and 1986, respectively.
In 1987, he was appointed Visiting Associate Professor with the University of Wisconsin, Madison, WI, USA, where he developed the DAB converter. In 1988, he joined the GE Corporate Research and Development Center, Schenectady, NY, USA. In 1994, he joined Silicon Power Corporation (formerly GESPCO) as Vice President Technology, developing the world’s first medium-voltage static transfer switch. Since 1996, he has been a Professor with RWTH Aachen University, Aachen, Germany, where he leads the Institute for Power Electronics and Electrical Drives. In 2006, he was appointed the Director of the E.ON Energy Research Center, RWTH Aachen University. Since 2014, he has been the Director of the German Federal Government BMBF Flexible Electrical Networks Research CAMPUS. He has authored or coauthored more than 800 technical papers and is holder of more than 60 patents.
Dr. De Doncker was the recipient of the IAS Outstanding Achievements Award, the 2013 Newell Power Electronics IEEE Technical Field Award, the 2014 IEEE PELS Harry A. Owen Outstanding Service Award, the IEEE Medal in Power Engineering in 2020, and the Doctor Honoris Causa degree of TU Riga, Latvia. In 2015, he was awarded Fellow status at RWTH Aachen University. In 2016, he became a Member of the German Academy of Science and Technology (ACATECH).
\end{IEEEbiographynophoto}

\begin{IEEEbiographynophoto}{Heike Vallery}
graduated with honors from RWTH Aachen University in 2004, with a Dipl.-Ing. degree in Mechanical Engineering. In 2009, she earned her Dr.-Ing. from the Technische Universität München and then continued her academic career at ETH Zürich, Khalifa University, and TU Delft. Today, she is a full professor at RWTH Aachen University and TU Delft, and also holds an honorary professorship at Erasmus MC in Rotterdam.  Her main research interests are in design and control of robotic systems. Heike Vallery received numerous fellowships and awards, such as the 1st prize of the euRobotics Technology Transfer Award 2014, and recently an Alexander-von-Humboldt professorship to join RWTH Aachen University.
\end{IEEEbiographynophoto}

\begin{IEEEbiographynophoto}{Maximilian Basler}
graduated with honors from the University of Duisburg-Essen in 2018 with a M. Sc. in mechanical engineering, specializing in mechatronics. He subsequently joined the Institute of Automatic Control at RWTH Aachen University, focusing his research on hardware-in-the-loop wind turbine testing, and earned his Dr.-Ing. in 2024. In 2020, he took over leadership of the Drive and Wind Energy Systems research group. Since 2023, as Head of Energy Systems at the Institute of Automatic Control, he has been responsible for all activities related to energy systems, including key research areas such as wind energy, grid integration, hybrid energy systems, and hydrogen technologies.
\end{IEEEbiographynophoto}

\end{document}